\documentclass[aps,prl,twocolumn,superscriptaddress,showpacs]{revtex4}
\usepackage{amsmath}
\usepackage{dcolumn}
\usepackage{epsfig}
\usepackage{graphicx}
\usepackage{latexsym}

\def\BISE{Bi$_2$Se$_3$}
\def\BITE{Bi$_2$Te$_3$}
\def\BISB{Bi$_{1-x}$ Sb$_x$}
\begin{document}



\title{STM imaging of impurity resonances on \BISE  }

\author{Zhanybek Alpichshev}
\affiliation{Stanford Institute for Materials and Energy Sciences, SLAC National Accelerator Laboratory, 2575 Sand Hill Road, Menlo Park, CA 94025}
\affiliation{Geballe Laboratory for Advanced Materials, Stanford University, Stanford, CA, 94305}
\affiliation{Department of Physics, Stanford University, Stanford, CA 94305}
\author{Rudro R. Biswas}
\affiliation{Department of Physics, Harvard University, Cambridge MA 02138}
\author{Alexander V. Balatsky}
\affiliation{Theoretical Division, Los Alamos National Laboratory, Los Alamos, NM 87545}
\affiliation{Center for Integrated Nanotechnologies, Los Alamos National Laboratory, Los Alamos, NM 87545}
\author{J. G. Analytis}
\affiliation{Stanford Institute for Materials and Energy Sciences, SLAC National Accelerator Laboratory, 2575 Sand Hill Road, Menlo Park, CA 94025}
\affiliation{Geballe Laboratory for Advanced Materials, Stanford University, Stanford, CA, 94305}
\author{J.-H. Chu}
\affiliation{Stanford Institute for Materials and Energy Sciences, SLAC National Accelerator Laboratory, 2575 Sand Hill Road, Menlo Park, CA 94025}
\affiliation{Geballe Laboratory for Advanced Materials, Stanford University, Stanford, CA, 94305}
\author{I.R. Fisher}
\affiliation{Stanford Institute for Materials and Energy Sciences, SLAC National Accelerator Laboratory, 2575 Sand Hill Road, Menlo Park, CA 94025}
\affiliation{Geballe Laboratory for Advanced Materials, Stanford University, Stanford, CA, 94305}
\affiliation{Department of Applied Physics, Stanford University, Stanford, CA 94305}
\author{A. Kapitulnik}
\affiliation{Stanford Institute for Materials and Energy Sciences, SLAC National Accelerator Laboratory, 2575 Sand Hill Road, Menlo Park, CA 94025}
\affiliation{Geballe Laboratory for Advanced Materials, Stanford University, Stanford, CA, 94305}
\affiliation{Department of Physics, Stanford University, Stanford, CA 94305}
\affiliation{Department of Applied Physics, Stanford University, Stanford, CA 94305}


\date{\today}

\begin{abstract}
In this paper we present detailed study of the density of states near defects in \BISE. In particular, we present data on the commonly found triangular defects in this system. While we do not find any measurable quasiparticle scattering interference effects, we do find localized resonances, which can be well fitted by theory  \cite{biswas} once the potential is taken to be extended to properly account for the observed defects. The data together with the fits confirm that while the local density of states around the Dirac point of the electronic spectrum at the surface is significantly disrupted near the impurity by the creation of low-energy resonance state, the Dirac point is not locally destroyed. We discuss our results in terms of the expected protected surface state of topological insulators.  
\end{abstract}


\maketitle

Probing the interaction of quasiparticles with impurities and defects has proven to be a powerful tool for the study of electronic structure in solids.  First, quasiparticle scattering from defects and impurities give rise to interference effects  of electron-waves associated with bands dispersion of the solid. In addition, impurities and defects can also result in bound states that induce localized resonances. For surface state bands (SSB), both phenomena are most suitably studied using scanning tunneling microscopy (STM) and spectroscopy (STS) \cite{crommie,hasegawa,sprunger}. In the case of topological insulators (TI) \cite{qz,hk}, probing quasiparticle scattering interference (QPI) and resonance scattering phenomena are expected to be particularly revealing, given that our focus is on the unusual properties expected from the SSB of these  materials. 

Much of the recent STM work on impurity and defect scattering in TI systems concentrated on   \BISB~\cite{roushan} and \BITE~\cite{alpichshev,zhang}. Indeed, within the main part of the  SSB, exhibiting convex contour, oscillations were shown to be strongly damped, supporting the hypothesis of topological protection.   However, oscillations do appear in the upper part of the SSB exhibiting concave contour \cite{alpichshev}, which disperse with a particular nesting wave-vector corresponding to the allowed spin states of the warped  contour \cite{fu1}. In contrast,  the cross-sectional contours of the SSB in \BISE~are invariably convex, and together with weak scattering amplitude for non time-reversed states could explain the lack of QPI, despite the observation of local impurities and defects in this system  \cite{urazhdin2,hor}. However, these defects do affect the local surface potential and therefore should also affect the Local Density of Sates (LDOS), exhibiting resonance phenomena similar to d-wave superconductors \cite{balatskyRMP} and graphene \cite{udega}.

In this paper we present a detailed study of  defects-induced resonance states on the surface of the 3D topological insulator \BISE.   Different defects are studied, all showing evidence for the creation of low-energy resonance states which vary in strength but do not locally destroy the Dirac point. We further show that the main features of the data are captured by  a simple extension of theoretical analysis of point-impurity scattering in the presence of Dirac dispersion \cite{biswas}, thus stressing the robustness of the SSB. In particular we present high resolution topography and spectroscopy studies of resonance states associated with the frequently observed triangular defects, while also making a direct connection between their occurrence and the samples bulk doping.

Single crystals of \BISE~were grown by slow cooling a binary melt as described elsewhere \cite{analytis1}, yielding samples with typical dimensions of $2 \times 1 \times 0.1$ mm$^3$.  The trigonal c-axis is perpendicular to the cleavage plane of the crystals. Low magnetic field Hall measurements of similar undoped samples indicated n-type doping  of $\sim 1\times 10^{19}$ carriers/cm$^3$ which is a result of Se deficiency \cite{analytis1,horak}.  In this paper we also show results on Sb-doped samples. These lower carrier density samples (about 100 times lower), with much reduced triangular defects, were obtained by slow cooling a ternary melt containing progressively more Sb. Although Sb is isovalent with Bi, its substitution acts to control the defect density in the bulk crystals, reducing the bulk carrier density \cite{lostak}. 

 \begin{figure}[h]
\includegraphics[width=1.0 \columnwidth]{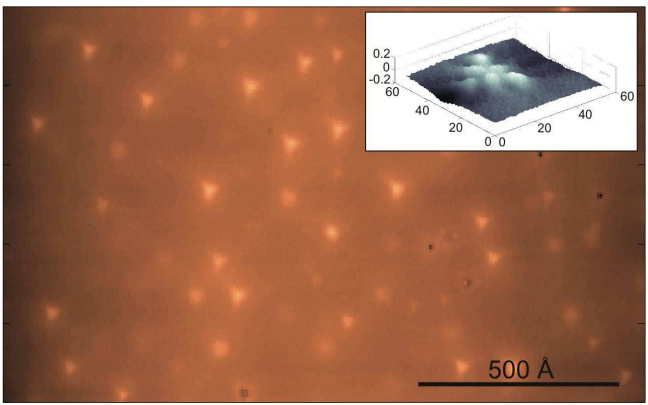}
\caption{1600$\times$1000\AA$^2$ topograph of the surface of undoped \BISE. Most  impurities are triangular (typical one marked), pointing in the same direction, and are a consequence of a sub-surface Se vacancies  \cite{urazhdin2,hor}. Inset shows a three dimensional rendering of a triangular defect.   }
\label{topo}
\end{figure}

Samples were cleaved in vacuum of better than $5\times10^{-10}$ Torr, and quickly lowered to the $\sim 8$ K section of the microscope, where cryopumping ensures that the surface remains free from adsorbates for weeks or longer. Topography scans were taken at several bias voltages and setpoint currents (typically +100mV and 60pA).  Figure~\ref{topo} shows a $1600 \times 1000 \AA^2$ topograph of the unoccupied states at the surface of cleaved undoped \BISE~ crystal.  Similar to previous studies of undoped \BISE~crystals \cite{urazhdin2,hor}, we observe  a large concentration of triangular defects,  uniquely oriented, with similar size but different brightness. High resolution topographs of the brightest triangular defects (see inset of Fig.~\ref{topo}) revealed a structure of six ``bumps," each about $\sim0.1\AA$ high, separated by a distance of 8.2$\pm 0.2\AA$, corresponding to twice the hexagonal lattice constant of \BISE. Less intense triangles have the same spatial structure but weaker in strength. Assuming Se vacancies as the main source of defects in our samples \cite{analytis1}, we can interpret the triangular structures as a response to the Se vacancies in different layers beneath the surface. As we will argue below, this interpretation is further supported by an earlier study of Urazhdin {\it et al.} \cite{urazhdin1}, where excess Bi was introduced to \BISE~crystals, producing clover-shaped defects in the occupied-states topographs, with the same dimensions as the triangular defects. Detailed calculations of these defects \cite{urazhdin1} suggested that they result from antisite doping where a Bi atom replaces a Se atom in the bottom Se layer of the top quintuple stack. The fact that resonance states can be identified in that paper (see Fig.~3a in \cite{urazhdin1}), with opposite charge compared to our data below, further supports our proposal. Counting $\sim 60$ triangular defects in this topograph, and assuming that they all come from within the first quintuple layer, each donating two electrons per defects, we find a bulk carrier density of $\sim 6 \times 10^{18}$ carriers/cm$^3$ similar to the one deduced from Hall effect. The much reduced triangular defects found for the Sb-doped samples, for which also the bulk carrier density is low, supports this scenario. 

  \begin{figure}[h]
\includegraphics[width=1.0 \columnwidth]{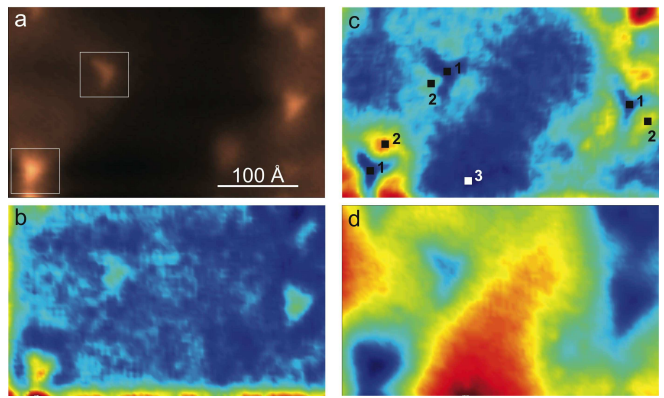}
\caption{A topograph (a) , and selected energy maps emphasizing two triangular impurities with different intensity,  while both are the same size,  bright triangle corresponds to a stronger protrusion than the  faint one.  LDOS at 90 meV (b), $-$180 meV (c) (points 1, 2, and 3 mark the center of defect, side-resonance, and background respectively) , and $-$400 meV (d) are also shown.  }
\label{mapst}
\end{figure}

Figure~\ref{mapst} depicts a topograph of a region containing several triangular defects, together with  LDOS at three energies: close to the bulk conduction band (BCB) (Fig.~\ref{mapst}b), in the middle of the exposed Dirac SSB (Fig.~\ref{mapst}c), and below the Dirac point close to the bulk valence band (BVB) (Fig.~\ref{mapst}d). In analyzing these maps (and many similar maps we studied), we concentrate on two triangular defects with different brightness, corresponding to a different amount of protrusion.  The first thing to notice that while LDOS in the presence of the BCB is clearly enhanced on the defect site (Fig.~\ref{mapst}b), the LDOS in the presence of the BVB is suppressed (Fig.~\ref{mapst}d). This  indicates that the defect site is positively charged, which again is confirmed by the fact that our undoped \BISE~is n-type.  However, the most interesting behavior is found within the exposed Dirac-SSB. As is seen in Fig.~\ref{mapst}c,  three clear LDOS peaks around the triangular defects are formed,  whose intensity strongly depend on energy. Although the lack of QPI implies a suppression of backscattering, the residual scattering of other momentum-transfer wave-vectors add up to a give rise to a resonance state \cite{biswas}.  

   \begin{figure}[h]
\includegraphics[width=1.0 \columnwidth]{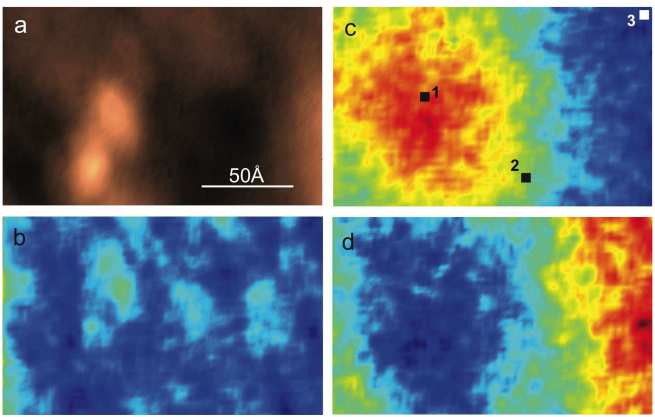}
\caption{A topograph (a) , and selected energy maps emphasizing an extended impurity in Sb-doped \BISE. LDOS at  75 meV (b), $-$225 meV (c) (relevant to Fig.~\ref{fits}, points 1, 2, and 3 mark the center of the defect, the edge of its extent, and background respectively)  , and $-$400 meV (d) are shown. }
\label{mapss}
\end{figure}

The Sb-doped \BISE~ samples show reduced density of triangular defects, but exhibit other types of defects. Fig.~\ref{mapss} shows a topograph of a typical defect (presumably Sb-cluster) and LDOS maps of the same region. Here again, for energies overlapping with the BCB and BVB the LDOS at the defect site and around it is depressed, while a strong peak, now spread at and around the impurity, appear in the exposed region of Dirac-SSB (Fig.~\ref{mapss}c).

\begin{figure}[h]
\includegraphics[width=1.0 \columnwidth]{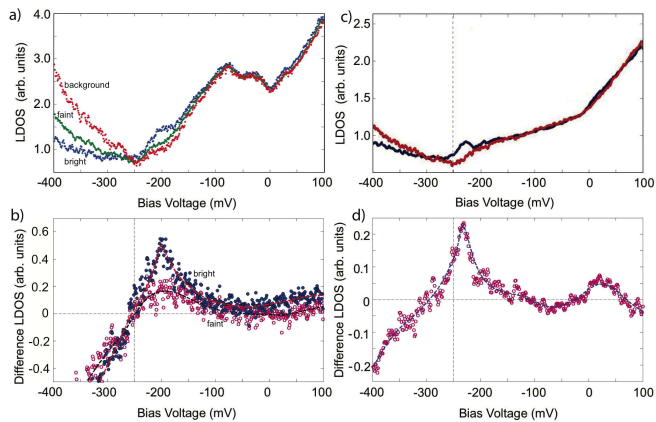}
\caption{Emergence of a resonance near the Dirac point for two different impurities;  a) Triangular defects corresponding to the ``bright" and ``faint"  of Fig.~\ref{mapst}, with the difference LDOS for both defects shown in (b);  c) Extended defect in Sb-doped \BISE ~(Fig.~\ref{mapss}), and the difference LDOS shown in (d). Dashed lines mark the location of the Dirac point.}
\label{resonance}
\end{figure}

Figure~\ref{resonance}a shows LDOS on the triangular impurities (``bright" and ``faint"), with a LDOS curve away from impurities shown for reference \cite{residual}.   Fig.~\ref{resonance}b shows the difference in LDOS between those curves, showing a broad resonance peak that is located at positive bias as compared to the Dirac point. Both resonances are rather similar in shape, and we emphasize here that similar shape and width have been measured for many triangular defects with a variety of ``brightness" on several different samples. Fig.~\ref{resonance}c shows LDOS on top of the defect shown in Fig.~\ref{mapss} on a Sb-doped crystal, together with the LDOS away from the impurity. Fig.~\ref{resonance}d shows the difference curve for these two spectra, resulting in a much sharper and more pronounced resonance.

\begin{figure}[h]
\includegraphics[width=1.0 \columnwidth]{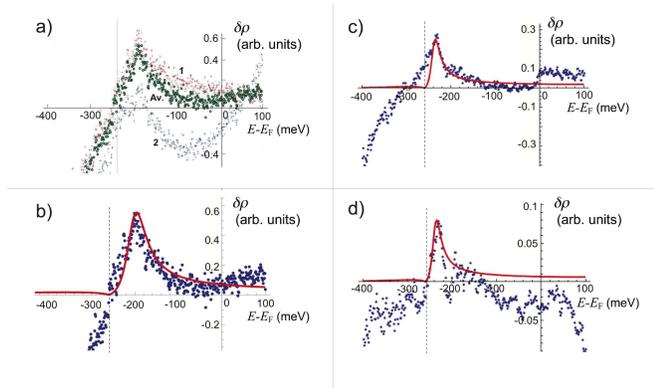}
\caption{Theoretical fits to resonance amplitudes.  a) raw LDOS difference data for a``bright" triangle where (1) represents the middle of the triangle and (2) represents the side peaks (see Fig.~\ref{mapst}c). The averaged data is marked with Av.; b) Averaged data from (a), fitted to theory (see text) with $V_0=3.17$ eV, $R=15 \AA$;  c) Fit to extended defect on an Sb-doped sample, at center of defect  (point (1) in Fig.~\ref{mapss}) with $V_0=0.71$ eV, $R=64 \AA$; d) Same parameters as for (c), but away from center of defect (point (2) in Fig.~\ref{mapss}) (see text). Dashed lines mark the location of the Dirac point.}
\label{fits}
\end{figure}

The general problem of a impurity-potential scattering on the surface of a TI has  been recently studied by Biswas and Balatsky \cite{biswas}. Considering  a two-dimensional Dirac gas interacting with a $\delta$-function-potential, they found that a resonance state is formed around any localized defect; the Dirac point is not destroyed by the scattering; and LDOS around the impurity falls of as $1/r^2$. Furthermore, the shape of the resonance is not Lorentzian and its width is determined by the strength of the potential which also sets the position of the resonance on the energy axis. The stronger the potential is the closer the peak gets to the Dirac point and the narrower it gets. While neither the triangular defects nor  the Sb-cluster, can be approximated by a $\delta$-function potential, we assume that the net effect on the SSB may still be similar, especially if we look at the LDOS slightly away from the center of the defect. However, to account for the extended nature of the defect we modify the calculations of ref.~\cite{biswas}, but with a step-potential replacing the local $\delta$-function potential. The step-potential is parametrized with a height $V=V_0$, extending over a radius $R<R_0$,  and zero everywhere else.  Using a step-potential, the strength of the impurity is therefore characterized by the product $V_0R_0^2$, rather than the potential height itself.  The extent of the potential ($R_0$) is further checked against the topography and the LDOS in the exposed part of the Dirac band.  In particular, since the center of the triangular impurity shows suppressed LDOS, the ``on-site" LDOS for these impurities are calculated as an average between the LDOS at the center and the LDOS at the three peaks surrounding it (shown in Fig.~\ref{fits}a).   Fig.~\ref{fits}b-c shows typical fits to the measured LDOS difference for both, triangular and Sb-cluster defects. 
 
The first thing that we notice is that the quality of the fit is visibly better for a triangular defect (Fig.~\ref{fits}b compared to the extended defect on the surface of an Sb-doped sample (Fig.~\ref{fits}d). We believe that this is due to the fact that the triangular defects are more compact in space and are therefore better described as a local perturbation. The Sb-cluster on the other hand is a significantly extended object whose effect might not be described as a simple potential interaction with SSB. For example, in a recent study of an extended step on the surface of Bi$_2$Te$_3$ we have shown that a ``bound-state" is formed as a consequence of interference in the bulk due to the finite penetration depth of the SSB \cite{theory}.  Specifically, an important feature of the theory is the protection of the Dirac point against the perturbation due to the impurity. Here, as the bare LDOS (far away from any impurity) is subtracted from the LDOS that is affected by the impurity, and assuming that the background incoherent LDOS is uniform in space and only weakly energy dependent, the zero LDOS should occur at the Dirac point. While this feature is satisfied for the triangle impurities, the Sb-cluster impurity shows excess scattering near the Dirac point, shifting the zero of the difference-LDOS ($\delta \rho$) by about 30 meV. However, as is seen in Fig.~\ref{fits}d away from the center of the defect the fit becomes much better, including the location of the Dirac point (zero crossing).  Therefore, the overall fit of our model to the extended defect data captures the shape of the resonance, and the fact that being a strong impurity it is sharper and peaks closer to the Dirac point. 

Focusing on the triangular impurities, we already mentioned that they appear with a variety of strengths related to their different protrusion heights (see Fig.~\ref{topo}.)  A surprising result, which is revealed when we examine the LDOS difference of the ``bright" and ``faint" triangular defects, is that they have the same scattering strength as both positions of the associated resonances and their peaks widths are the same within the error-bars. This is in a good agreement with the suggestion that all triangle impurities have similar origin.

\begin{figure}[h]
\includegraphics[width=1.0 \columnwidth]{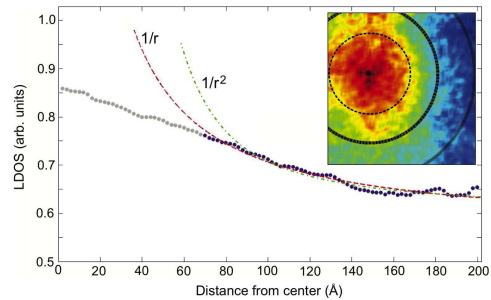}
\caption{LDOS as a function of distance from the center of the impurity. Dashed line is a fit to $1/r$ for the dark points (after $\sim 65\AA$) and dash-dotted line is a fit to $1/r^2$. Inset shows the contours of decay of the LDOS around the impurity center ($\bullet$) at 65$\AA$, 100$\AA$, and 150$\AA$. }
\label{distance}
\end{figure}

Finally we show in  Fig.~\ref{distance}  the dependence of the LDOS at the energy of the resonance of the broad scatterer (Sb-cluster) on the distance away from the impurity center. Fitting the decay as a function of distance, we find it hard to determine the exact power law decay. In Fig.~\ref{distance} we show an averaged decay for the Sb-cluster defect, fitted with both, $1/r$ and $1/r^2$. While the $1/r$ decay fits the data better all the way to the extent of the defect, the two power laws, and in fact any power law in between may still be a good description of the data, especially considering the large characteristic length for \BISE~ surface state ($\lambda \sim 30 \AA$ \cite{zhang}). In particular the data rules out an exponential decay of the LDOS resonance with $\lambda$, which is another confirmation of the two-dimensional nature of the band giving rise to the resonance state.
 
In conclusion, we identified the resonance states associated with the triangular impurities in \BISE. We found that independent of their observed intensity, these impurities induce resonances of similar shape and strength, thus can be considered as a characteristic of \BISE~(and possibly also \BITE) spectroscopy. We further showed that other type of defects, extended in nature, also give rise to resonant states. Comparing the various defect resonances to each other and with theory, we conclude that a simple scattering mechanism in the presence of a Dirac-type 2D quasiparticles give a full account to the observed data. While theoretical curves fitted the data very well for the case of the triangular defects,  we find that for stronger, more extended defects,  some modifications are needed that may include the finite penetration depth of the surface state.

\acknowledgments
This work was supported by the Center on Functional Engineered Nano Architectonics (FENA) and the Department of Energy Grant  DE-AC02-76SF00515. Work at Los Alamos was partially supported  by The Center for Integrated Nanotechnologies, and in part by the LDRD through the LANL under Contract DE-AC52-06NA25396 and UCOP TR-027.


\begin{thebibliography}{99}
\bibitem{residual}
Note that for \BISE~the Dirac point is exposed, and thus one would expect a linear LDOS as a function of energy all the way to the Dirac point where zero LDOS should be observed. However, a large background LDOS, presumably incoherent in nature, shifts the Dirac point to a large LDOS value, a fact that needs to be taken into account when any data involved the SSB is analyzed.
\end{thebibliography}
\end{document}